# Online Experimentation with Surrogate Metrics: Guidelines and a Case Study

WEITAO DUAN, LinkedIn Corporation
SHAN BA, LinkedIn Corporation
CHUNZHE ZHANG, LinkedIn Corporation

A/B tests have been widely adopted across industries as the golden rule that guides decision making. However, the long-term true north metrics we ultimately want to drive through A/B test may take a long time to mature. In these situations, a surrogate metric which predicts the long-term metric is often used instead to conclude whether the treatment is effective. However, because the surrogate rarely predicts the true north perfectly, a regular A/B test based on surrogate metrics tends to have high false positive rate and the treatment variant deemed favorable from the test may not be the winning one. In this paper, we discuss how to adjust the A/B testing comparison to ensure experiment results are trustworthy. We also provide practical guidelines on the choice of good surrogate metrics. To provide a concrete example of how to leverage surrogate metrics for fast decision making, we present a case study on developing and evaluating the predicted confirmed hire surrogate metric in LinkedIn job marketplace.

## 1 INTRODUCTION

With the fast development of Internet, an unprecedented amount of new ideas are generated constantly in online services and applications. As a result, there is a growing need for quickly evaluating ideas and understanding the results. A/B tests (or controlled experiments, split tests) have been widely adopted in the online world as the golden rule for decision making and the driving forces for innovation. Many technology companies, such as Microsoft, Google, Facebook, LinkedIn, Uber, Netflix and Twitter, have in-house experimentation platforms, where experiments are run at large scale with marginal costs [2, 10, 12, 19, 25, 27–29]. From user-interface changes to back-end algorithms and infrastructure, from software developers to product managers to data scientists, A/B tests help make data-driven decisions and innovate new product ideas.

How fast we innovate can be limited by how quickly we can conclude the experiment result. At LinkedIn, most experiments go through a "ramp-up" process. Because we do not know a priori how the feature performs, we start by ramping the feature to a small percentage of users and then check if their business as well as operational metrics are good. After we gain more confidence on the performance, we ramp to more users and repeat the measurement until we increase the traffic of the new feature to 100%. LinkedIn has built up a ramping and experimentation framework that can effectively balance Speed, Quality and Risk (SQR) [30]. During low ramp stage, the goal is to mitigate risk, and quickly dial up the ramp when metrics are good or neutral until reaching maximum power ramp. At maximum power ramp (50% ramp in the case of one treatment and one control), we spend enough time to collect experiment results and watch out for feature burn-ins as the goal is to properly measure the impact. The SQR framework has been regarded as *the* standard for how people ramp experiments at LinkedIn and is deeply integrated into our experiment ecosystem.

This works well for online experiments where the treatment effects on the metrics of interest are instantaneous or can be observed in a relatively short period of time. For instance, the improvement on web page load time from a new backend system can be observed and measured hours even minutes after the ramp. The improved sign-up conversation from a new ad campaign can be measured the following hours or days after the launch.

However, in many scenarios, it takes a long time to measure the treatment impacts precisely. For instance, in subscription-based products (such as LinkedIn Premium Member Subscription), we want to evaluate how the treatment impacts a member's premium subscription lifetime value (LTV). A member's LTV usually can only be partially observed until the member cancels the subscription. Similarly, in LinkedIn Job Marketplace, our goal is to improve the effectiveness of a new feature or a new recommendation algorithm in helping members land a job. Although the new feature or algorithm usually has quick impacts on job matching, interaction and application results, its downstream impact on hires will not be observed for several months in the future. In marketing and e-commerce field, we want to optimize for user's conversion. However, conversions are often lagged behind impressions and clicks, and can take days, weeks or even months to realize.

One may suggest that we just run the experiment for longer. There are several reasons why we cannot simply let such an experiment run for months until its treatment impact on the long-term true north metric can be fully observed. First, since a new feature may be bad and hurt member experience, doing so puts our member experience at risk and violates LinkedIn's "member-first" principle. Secondly, running for months would substantially slow down our experiment velocity and cannot keep up with our innovation speed. Thirdly, keeping an experiment running for a long time greatly increases the chance of having unexpected interactions or incompatible issues with other experiments. Last but not least, keeping a large number of experiments running for months poses not only heavy computation burden to the experiment platform, but also management burden to the product managers as the product experience becomes fragmented.

In order to assess the treatment effects in a timely manner for these applications, we often build a machine learning model to predict the long-term true north metric (i.e. LTV or hires) based on a few easier-to-observe predictors. The model prediction can be used as an approximation for the true north and is often referred to as a *surrogate metric*, *proxy metric* or *surrogate*. In practice, the surrogate metrics are often used directly as if they were the ground truth in a regular A/B test to drive critical business decisions. After a surrogate metric is determined, people seldom take into consideration the errors in its underlying predictive models for approximating the true north. In this paper, we show that running regular A/B tests on surrogate metrics tends to inflate the Type I error. Without further taking in account the uncertainty of surrogate metrics, the A/B test could give misleading conclusions.

Key contributions of this paper include:

- We show why direct comparison on surrogate metrics will inflate the experiment Type-I error.
- We discuss, when using surrogate metrics, how to adjust the A/B testing comparison to ensure experiment results are trustworthy.
- We provide guidelines for choosing the right surrogate metrics from both practical and theoretical perspectives.
- We construct the work flow of leveraging surrogate metrics for quick decision making
- We present a case study on developing and evaluating the predicted confirmed hire surrogate metric in LinkedIn job marketplace experimentation.

The rest of this paper is organized as follows. In Section 2, we review the foundation of controlled experiments. Section 3 discusses the practical guidelines and theoretical conditions on the choice of good surrogate metrics. In

Section 4, we dive into the challenges of using surrogate metrics in experiment setting and propose our solutions. We then walk our readers through a real world example at LinkedIn and demonstrate how we leverage surrogate metrics in experiment decision making in Section 5. Section 6 concludes the paper with future work.

## 2 REVIEW ON CONTROLLED EXPERIMENTS

In this section, we give a brief review on the evolution of controlled experiment techniques with a focus on experiment ramping and evaluation. The foundation of experimentation was introduced by Sir Ronald A. Fisher at the Rothamsted Agricultural Station in England in the 1920s with a focus on agriculture [3]. While the theory is simple, many researchers have studied and extended Fisher's work in textbooks and papers [26] and controlled experiment has gained its popularity beyond the original agricultural field [17, 18]. Experiment practitioners in many fields leverage the theory and conduct experiments to evaluate new ideas [7, 27]. In the technology industry, experimentation is adopted by many companies [2, 12, 13, 19, 25, 29]. Deployment and analysis of controlled experiments are done at large scale. This presents unique challenges and pitfalls. Many researchers and experiment practitioners have described the challenges, pitfalls and novel solutions in publications [4, 8, 9, 29].

One of the shared challenge, among many, is to be able to make decision fast and conclude early. There are mainly three fronts of work. The first tackles the problem at the design stage with novel experiment designs (such as overlapping experiments) to achieve high power [27]. The second front aims at improving the analysis stage and removing noise by either incorporating historical user behaviors or experiment results [5, 7] .The third front leverages machine learning models to build surrogate metrics for estimating the long-term impact [1].

Before we dive into the how we address the challenge and pitfalls of evaluating long-term impact while iterating fast, we want to review the set up and notation for controlled experiments and lay the foundation for the following sections.

Suppose we have one treatment feature $T$ and one control experience $C$, the metric of interest for user $i$ is $Y_i$ and the assignment for user $i$ is $W_i$, where

$$W_i = \begin{cases} 1 & \text{if user } i \text{ is in treatment group} \\ 0 & \text{if user } i \text{ is in control group} \end{cases} \quad (1)$$

Following Rubin Causal Model or the potential outcome framework set up [14, 15, 24], each unit's potential outcome is defined as a function of the entire assignment vector $\mathbf{W} \in \{0, 1\}^N$ with $N$ of units to treatment buckets: $Y_i(\mathbf{W})$. If Stable Unit Treatment Value Assumption (SUTVA) holds, the realized outcome and the potential outcome have the following relationship:

$$Y_i = \begin{cases} Y_i(0) & \text{if } W_i = 0 \\ Y_i(1) & \text{if } W_i = 1 \end{cases} \quad (2)$$

Suppose there are $N_C$ units in the control group and $N_T$ units in the treatment group, the Average Treatment Effect (ATE) is given by:

$$\mu_Y = \frac{1}{N_T} \sum_{i \in T} Y_i(1) - \frac{1}{N_C} \sum_{i \in C} Y_i(0) \quad (3)$$

## 3 SURROGATE METRICS

### 3.1 Average Treatment Effect from Surrogate Metrics

Suppose that we have equal sized treatment and control group. Among $2n$ total units, $n$ units are randomly assigned to the treatment group $T$ and the rest to the control group $C$. The average treatment effect (ATE) $\mu_Y$ based on the long-term metric $Y_i$ under SUTVA is

$$\mu_Y = \frac{1}{n}[\sum_{i \in T} Y_i W_i - \sum_{i \in C} Y_i(1 - W_i)] \tag{4}$$

Because $Y_i$ is not observable in a short period of time, we resort to some observable predictors $\mathbf{X_i}$, and fit a machine learning model to predict the true north metric $Y_i$ of a unit $i$. The model prediction can then be used as an approximate for the true north metric, and we call it the surrogate metric $S_i$.

Specifically, we have

$$Y_i = S_i + \epsilon_i = f(\mathbf{X_i}) + \epsilon_i, \text{ for } i = 1, 2, ..., 2n \tag{5}$$

where $\epsilon_i$'s represent the errors from using surrogate metrics to approximate the long term outcomes. Based on surrogate metrics, the ATE is estimated by

$$\mu_S = \frac{1}{n}[\sum_{i \in T} S_i W_i - \sum_{i \in C} S_i(1 - W_i)] \tag{6}$$

Comparing equation (6) with (4), if errors $\epsilon_i$ are zeros (i.e. $S_i$ can perfectly predict the true north metric $Y_i$), the comparison on the surrogate metric is equivalent to that on the true north. However, this is not the case for most predictive models.

### 3.2 Guidelines for Choosing the Right Surrogate Metrics

In practice, there are often more than one ways to build a predictive model and approximate the long-term true north metric $Y_i$. [6, 20, 21] all have studied the topic of choosing the right surrogate metrics. Here we summarize a few practical guidelines for picking the right surrogate metrics.

*High predictive power on the true north.* As shown in (5), an ideal surrogate metric should be unbiased $E(\epsilon_i) = 0$ and has a high correlation with the true-north metric (i.e. $corr(S_i, Y_i)$ close to 1).

*Focusing on metrics we can change and measure in the short term.* The goal of constructing a surrogate metric is to make decision sooner. When we pick a surrogate metric, we need to think about what levers we can pull and what metrics we can move in the short term. Using a surrogate metric that is equally or almost equally hard to move or observe in the short term does not help conclude the result faster. We can leverage historical experiments to help us find levers and short-term metrics that are easier to move and have strong predictive power.

*Customization for different treatment features.* It is often ignored that the choice of surrogate metrics actually has to do with the mechanism of the treatment feature. For instance, we want to find a surrogate metric to predict long-term user session. If the treatment feature is implemented on the mobile application, picking a mobile-only metric, such as the number of page views on mobile, will likely give higher predictive power comparing to a desktop and mobile combined metric, even though both metrics potentially satisfy the statistical validity assumption in Section 3.3.

*Interpretability.* Many teams have tried to build machine learning models to create surrogate metrics and use them for decision making. For example, [22] uses a sequences of user interaction to predict search satisfaction. These metrics have greatly expanded the pool of candidate surrogate metrics and sparked research and discussion in this field. However, this is a fairly new field and there are some concerns with these machine learning models and the metric they create.

Especially, when the underlying machine models are complex and hard to understand, users often find these metrics untrustworthy because they cannot understand why the metrics move up or down. In this case, the adoption of such surrogate metrics is greatly discounted by the lack of understanding.

*Management overhead.* As mentioned before, a good surrogate metric may only work well for a subset of treatment features. We need to educate our users on the scenarios where the surrogate metrics tend to work well. Users tend to appreciate a simple "if-else" logic more ("if Scenario 1, then use Metric A; else use Metric B") and are more willing to adopt these metrics. If the logic is too complex, we need to either have this information ready on the experimentation platform at the time of concluding the result, or integrate the logic into the platform such that we automatically surface the right surrogate metric for different experiments. On the other hand, if the surrogate metric is powered by a machine learning model, the model may need to be constantly retrained. When the model is refreshed, the platform needs to respect the metric refresh and proactively backfill experiment result. Also, it may create confusion for digesting the result from an active experiment as the metric value may have changed from one day ago. Lastly, when the model is trained using the data from an actively experiment, the experiment and model can interact and introduce bias to the results.

## 3.3 Statistical Validity Requirement

Besides practical concerns, the surrogate metric needs to satisfy certain theoretical properties. Prentice [23] introduced the term "statistical surrogates". He laid out the formal requirement for a variable $S$ to be a statistical surrogate. Besides 1) treatment impacts the surrogate metric; 2) the surrogate metric impacts the long-term outcome, the validity assumption requires that conditional on the surrogate, the treatment and the final outcome are independent. In other words, for the set of all treatments in our consideration, there is a single pathway from treatment to the final outcome that goes through the surrogate metric. Therefore, once we know the surrogate metric, no other information is needed to determine the distribution of the final outcome. In mathematical notation, we have:

$$P(Y = y|S, W) = P(Y = y|S), \qquad (7)$$

where $S$ represents the surrogate metric. The only causal path is $W \to S \to Y$. It should be noted that even if the treatment had no direct impact on the long-term outcome $Y$, the confounding between the surrogate metric and the long-term outcome would invalidate the statistical surrogate assumption. [11, 16].

Prentice criteria, in practice, is very restrictive, because perfect surrogacy is unrealistic. It is almost impossible to find a perfect surrogate metric that satisfies all the conditions. While we cannot expect $P(Y = y|S, W)$ to equal $P(Y = y|S)$ exactly, these two quantities should not be too far apart for a good surrogate metric.

## 4 USING SURROGATE METRICS IN A/B TESTING

### 4.1 Errors from Direct Comparison on Surrogate Metrics

From (5), we can see that errors from using surrogate metric $S_i$ to approximate $Y_i$ are determined by $\epsilon_i$. Here we use a simple example to illustrate the pitfall of direct comparisons on surrogate metrics. Without loss of generosity, assume $\epsilon_i \sim N(0, \sigma^2)$ are i.i.d. and orthogonal to $f(\mathbf{X_i})$, and it is easy to see that

$$E(\mu_Y) = E(\mu_S) \qquad (8)$$

$$Var(\mu_Y) = Var(\mu_S) + \frac{2}{n}\sigma^2 \qquad (9)$$

In other words, the ATE based on the surrogate metric is unbiased, but the variance of the surrogate metric underestimates the variance of the true north. Without taking into account the the measurement errors of the surrogate metrics, the ATE on $\mu_Y$ tends to have more false positives. In the extreme case where the surrogate metric has zero predictive power on the true north, we should not trust any result based on the surrogate metric even if the $\mu_S$ is statistically significant.

Define the predicted R-squared as $R^2_{pred} = Var(\mu_S)/Var(\mu_Y)$. If Central Limit Theorem is applied, let $p_S$ and $p_Y$ denote the p-values on the surrogate metrics and the long term outcome had we observe it, we can show that the two-sided p-value from the surrogate metric underestimates the true p-value by

$$\begin{aligned}
\delta_p &= p_Y - p_S \\
&= 2\Phi\left(-\left|\frac{\mu_Y}{\sqrt{Var(\mu_Y)}}\right|\right) - 2\Phi\left(-\left|\frac{\mu_S}{\sqrt{Var(\mu_S)}}\right|\right) \\
&= 2\Phi\left(-\left|\frac{R_{pred}}{CV_S}\right|\right) - 2\Phi\left(-\left|\frac{1}{CV_S}\right|\right)
\end{aligned}$$

where

$$CV_S = \frac{\sqrt{Var(\mu_S)}}{\mu_S} \tag{10}$$

and $\Phi$ is the cumulative distribution function of a standard normal variable.

Figure 1 plots $p_S$ vs. $p_Y$ under different $R^2_{pred}$. As we can see, with a lower predictive power from the surrogate metric, the wider gap between the true north and surrogate p-values, hence, the higher chance of obtaining a false positive result. For example, with $R^2_{pred} = 0.85$, a p-value of 0.05 on the surrogate underestimates the true p-value by nearly 30%! (p-value on true north is 0.07)

To further illustrate how the Type I error is inflated in the comparison on surrogate metrics, we will walk our reader through a simulation study. Suppose the true north metric follows the relationship below:

$$Y_i = \frac{2}{3}exp(x_{i1}) - x_{i3}\,sin(x_{i2}) + x_{i2} \tag{11}$$

where $X_i = [x_{i1}, x_{i2}, x_{i3}]$.

For simplicity, we fit a linear model $S(X)$ as the surrogate metric with predicted R-squared $R^2_{pred} = 0.951$. Suppose in the treatment group, $x_{i1} \sim U(0, 1)$, $x_{i2} \sim U(0.14349, 1.14349)$ and $x_{i3} \sim U(0.15, 1.15)$ while $x_{ij} \sim U(0, 1)$ in the control group. Simulation code is available at https://anonymous.4open.science/r/b25ed0d0-5759-4a13-a65a-a02c80cbcb8a/. We can show that the averages of treatment and control group are equal. With 10,000 simulated samples, 560 comparisons on the surrogate metrics are statistically significant, which greatly exceed the expected 500 comparisons under $\alpha = 0.05$.

### 4.2 Quantifying Prediction Error of the Surrogate Metric

The previous subsection has shown the importance of incorporating error from the surrogate metric in A/B testing. Here we present two general approaches to assess the predictive power of surrogate metrics.

First, if we can have access to the underlying model in (5) that generates the surrogate metrics, we can output the corresponding prediction variances simultaneously. For most parametric statistical models, quantifying the prediction uncertainties should be quite straightforward. If certain non-parametric models or black-box models are used to predict the true north metric, we can leverage cross validation to quantify the prediction variance.

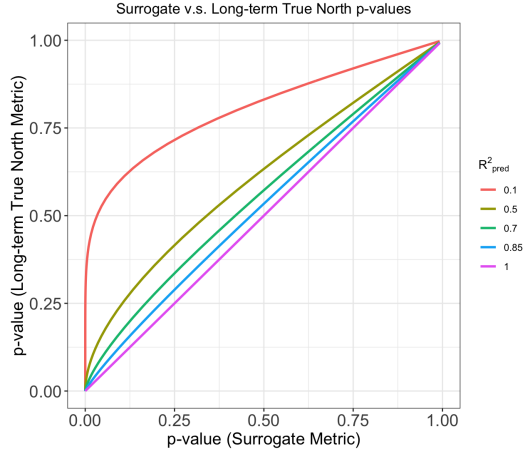

Fig. 1. p-values From Surrogate Metrics and Long-term True North Metrics under Various $R^2_{pred}$

Secondly, if we can only read the surrogate metrics but have no detailed knowledge about its underlying prediction model, a more general approach is to run back-tests based on the historic true north measurements and surrogate metrics. For example, suppose a long-term metric can only be observed six months after the event, we can take the historic surrogate metrics from half a year ago and compare them to the observed true north metrics to quantify the prediction errors from the surrogate model. This process can be repeated daily or in batch to generate a portfolio of historical prediction errors. We can then assess the predictive power from the surrogate model in this way.

### 4.3 Adjusting the Variance in A/B Testing

To control for Type-I error, we need to incorporate the predictive error $\sigma^2$ from surrogate metrics into the A/B comparison.

The adjusted t-statistic based on the surrogate metric can be written as

$$t_Y = \frac{\mu_Y}{\sqrt{Var(\mu_Y)}}$$
$$\approx \frac{\mu_S}{\sqrt{Var(\mu_S) + \frac{2}{n}\sigma^2}}$$

Clearly, if the measurement error variance were ignored (i.e. mistakenly setting $\sigma^2 = 0$), the p-value would be underestimated which leads to a high Type-I error.

On the other hand, however, as the predictive error of surrogate metric gets higher, the power of the A/B test becomes lower. In the extreme case, if $\sigma^2 \to \infty$ (i.e. predictions from the underlying model in ( 5 ) were useless), the t test would never declare any significance no matter how big the sample size is. As a result, when surrogate metrics are used in A/B testing, variance reduction techniques would be very helpful in maintaining the power of the test and improving the experiment velocity.

In the next section, we will dive into an important example of surrogate metrics at LinkedIn and how we leverage these principles for quick decision making.

## 5 CASE STUDY: SURROGATE METRIC FOR CONFIRMED HIRES AT LINKEDIN

In this section, we use an example in the LinkedIn hiring platform to demonstrate how to work with surrogate metrics. The same workflow can be extended to many other applications.

### 5.1 True North Metric for the Hiring Funnel on LinkedIn

The mission of LinkedIn's jobs marketplace is to help companies hire top talents and help people get jobs. To create an efficient job marketplace, we need to achieve three goals: 1) enable job seeker to better discover relevant job opportunities; 2) provide qualified job applications to job posters; 3) ensure that each job post receive sufficient number of job applications, but not too overwhelming. The team is constantly looking for opportunities to improve the hiring products with a true north metric called confirmed hires (CH). CH measures members who have found jobs with the help of LinkedIn products. By doing this, the team can better understand the value that LinkedIn product delivers to job posters and job seekers.

A major limitation of the CH metric is its long lag, because the CH calculation requires new hires updating their new job positions on their LinkedIn profiles. In the past, the success of product changes had to be measured by some top-of-funnel metrics, such as total job views and total job applications (see Figure 2) in A/B testing. However, an increase in top-of-funnel metrics is only aligned with the first goal but not the other two. The team had also developed other metrics such as job interactions (including profile view interactions and messaging interactions) to measure the two-way interest between job seekers and job posters. Although jobs interaction metric could provide some visibility into the second goal, it is only available for a subset of job segments. Moreover, different metrics may give mixed signals (e.g. job interactions go up while total applications go down), which makes the ramp decisions even more difficult.

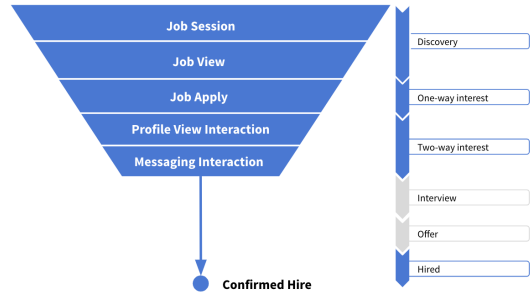

Fig. 2. Job Seeking Funnel

### 5.2 Building the Surrogate Metric

To tackle the above challenges, predicted confirmed hire (PCH) metric has been developed as a surrogate to get an early read on the true north metric (CH). The PCH model leverages job segments, application quality, and application distribution signals to predict each application's likelihood of becoming confirmed hires.

The PCH model can be described by

$$PCH = F(a_{jkqp}, b_{ji}, c_{ilqp}, d_{ij}, t), \tag{12}$$

where the notations have the following interpretations:

| | |
|---|---|
| $i$ | job seeker |
| $j$ | job post |
| $k$ | job application to a job post |
| $l$ | job application from a job seeker |
| $q$ | job segment |
| $p$ | application quality signal |
| $t$ | month when the application is submitted |
| $a_{jkqp}$ | indicator variable representing whether the $k$th application to job posting $j$ belongs to job segment $q$ and has quality signal $p$ |
| $b_{ji}$ | number of applications received for job posting $j$ after job seeker $i$ applies |
| $c_{ilqp}$ | indicator variable representing whether the $l$th application from job seeker $i$ belongs to job segment $q$ and has quality signal $p$ |
| $d_{ij}$ | number of applications already submitted by job seeker $i$ after applying to job posting $j$ |

The output of the PCH model is the predicted likelihood of the job application (by job seeker $i$ to post $j$) becoming a confirmed hire. All signals except job interactions in the PCH model have no lag while jobs interaction signals are available within a few days after application submission. Therefore, the overall lag of PCH is only a few days. Unlike the top-of-funnel metrics, PCH is sensitive to a wide range of application quality signals and application distribution signals. In other words, in addition to application volume, PCH favors better matching between job applicants and job posts, promotes early applications, and prevents any job from being overwhelmed with too many applications. PCH model has a closed form formula. Comprehensive sensitivity analyses have also been performed on the fitted model to understand how each factor impacts PCH. Considering that the goal of this paper is to demonstrate how to correctly measure surrogate metrics in online experimentation, we will focus on manifesting the process of using PCH in online experiments in the following sections.

### 5.3 Evaluating the Surrogate Metric

*5.3.1 Correlation between CH and PCH.* To check the predictive power of the surrogate metric, we compare PCH and our true north (CH) at the application level. We cut PCH into equally spaced buckets. For each PCH bucket, we calculate the average of CH for all applications within that bucket. The results are visualized in Figure 3. It shows that PCH and CH are highly correlated. The fitted line has a slope = 1.0157.

*5.3.2 Surrogacy Validity for PCH.* Although the above results show that PCH and CH are highly correlated, it is not sufficient to conclude that PCH is a statistically valid surrogate for CH. In order to leverage PCH in experimentation setting, we need to validate that, for the set of experiments we consider, PCH satisfies the surrogacy criteria and we can establish the causal path way $treatment \to PCH \to CH$.

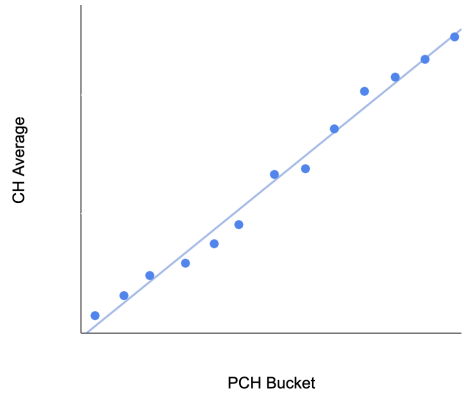

Fig. 3. PCH vs CH at job application level

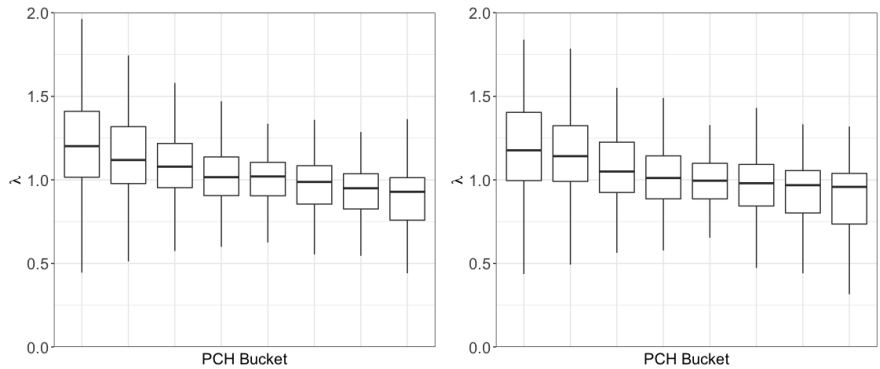

Fig. 4. $\lambda$ vs. PCH Bucket. Left: Control Group $W^k = 0$; Right: Treatment Group, $W^k = 1$

To carry out this study, we use experiments that aimed at improving the top of funnel (e.g. job applications) in the first quarter of 2019. These experiments range from job alert notifications that deliver new job posts on mobile devices, to user interface and relevance algorithm changes. The experiments were typically run for a few days to 2 weeks. Because PCH metric was not in production during that time, and we first back filled PCH for this period. Since most of Confirmed Hire can be observed in 6 months, at the time of evaluation, we can assume that we have observed both the surrogate (PCH) and the true north (CH) for the first quarter of 2019.

Recall from Section 3.3 that in order for PCH to satisfy the statistical validity assumption, the following has to be met:

$$P(CH = y|PCH, W^k) = P(CH = y|PCH), \tag{13}$$

where $W^k$ is the treatment assignment for experiment $k$ in the set of experiment we consider. $W^k = 1$ for treatment assignment, and $W^k = 0$ for control. Since CH is a binary variable, equation (13) can be simplified to

$$E(CH|PCH, W^k) = E(CH|PCH). \tag{14}$$

Since PCH is a continuous metric, $E(CH|PCH = x, W^k)$ can be highly influenced by a single observation in the experiment. To solve this, we bucketize PCH and compare $E(CH|PCH$ in bucket $j, W^k)$ with $E(CH|PCH$ in bucket $j)$. Let

$$\lambda_j^{W^k} = \frac{E(CH|PCH \text{ in bucket } j, W^k)}{E(CH|PCH \text{ in bucket } j)}. \quad (15)$$

Figure 4 plots $\lambda_j^{W^k}$ for all the experiments under consideration. We separate the cases of $W^k = 0$ and $W^k = 1$. From the plot, we can see that $\lambda_j^{W^k}$'s are not exactly one, but close. In practice, it is usually impossible to find a surrogate metric that satisfies $\lambda_j^{W^k} = 1$ for all the experiments under consideration. The box plot tells us that the surrogacy validity criteria are loosely satisfied.

We have verified the statistical surrogacy validity of PCH. However, we have not examined whether PCH can be moved in the experiment setting. First, we need to verify that a large portion of the new features are helpful in finding a job in the long-term, and they tested statistically significant on CH. Second, because we make decisions on PCH, we need to study whether the decisions based on PCH agree with CH.

*5.3.3 Estimating Prediction Variance.* Out of the 203 experiments in our study, we find out that on PCH, 30 experiments are statistically significant using alpha level $\alpha = 0.05$. Also, if we plot the t-statistic from the comparisons on PCH against CH on all 203 experiments, we see that the PCH and CH exhibit a strong linear relationship with $R^2 = 0.69$. This gives us a strong indication that PCH can help make early decisions on CH.

Now, we zoom into experiments that are deemed significant on PCH. We see that t-statistic on PCH tends to be more extreme and highly statistically significant, while t-statistic on CH from the same comparison is much smaller in magnitude (Figure 5). Recall from Section 4.1 that p-values on the surrogate metrics tend to much smaller than those on the true north when not incorporating the prediction error. Therefore, it is not surprising to see this relationship. Assuming we have $n$ units in treatment and control group respectively, we can recover the true error on CH with the following adjustment:

$$Var^{Adj}(\mu_{PCH}) = Var(\mu_{PCH}) + \frac{2}{n}\sigma^2 \approx Var(\mu_{CH}) \quad (16)$$

For simplicity, we assume the predicted mean squared error $\sigma^2$ for all members are the same and independent. We can estimate $\sigma^2$ by

$$\hat{\sigma}^2 = \frac{1}{N}|PCH_i - CH_i|^2, i = 1, 2, ..., N \quad (17)$$

during PCH model validation stage where $N$ is the total number of units in validation data set. In practice, the assumption of equal error may not be true. However, if our experiment population is similar to our training and validation set (which includes all job seekers on LinkedIn), $\hat{\sigma}^2$ can be used as a good estimate for PCH prediction error from the treatment or control group.

After incorporating the prediction error, we examined the relationship between $Var^{Adj}(\mu_{PCH})$ and $Var(\mu_{CH})$ as well as between adjusted PCH t-Statistics and CH t-Statistics. We zoom into experiments with significant impact on PCH. Figure 6 shows that $Var^{Adj}(\mu_{PCH})$ and $Var(\mu_{CH})$ match well. Figure 7 compares adjusted PCH t-Statistics and CH t-Statistics. As expected, they match well, especially more so on experiments with larger sample sizes (>1M). However, since the variances from adjusted PCH are now much higher, only 2 experiments are significant, compared to 30 before.

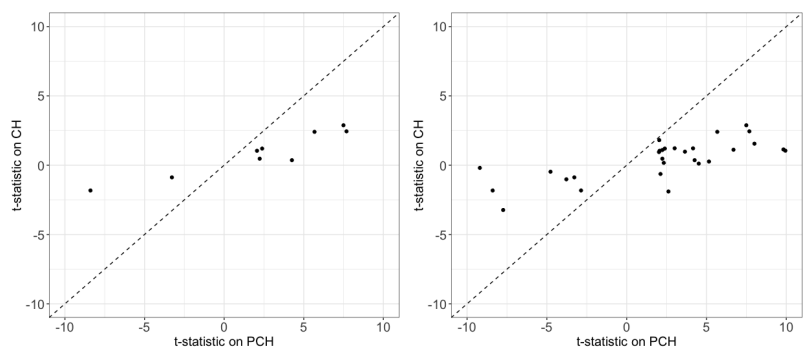

Fig. 5. Unadjusted PCH t-Statistic vs. CH t-Statistic. Left: Experiments with Sample Size > 1M in Both Treatment and Control; Right: Experiments with Sample Size > 10K in Both Treatment and Control. Plot Showing Experiments with Significant Impact on PCH)

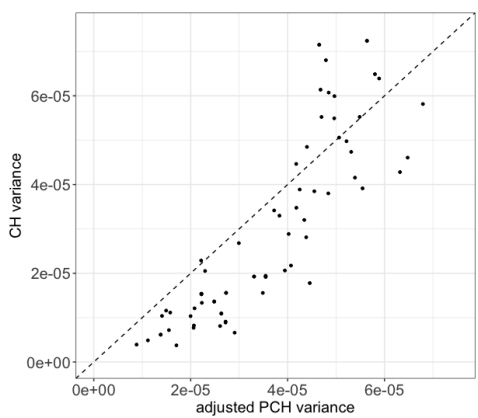

Fig. 6. Adjusted PCH variance vs. CH variance

### 5.4 Using Covariates for Variance Reduction

After incorporating the predictive error, the comparison on the surrogate metrics becomes noisier than the unadjusted. Many comparisons that were statistically significant before become inconclusive. The goal of introducing surrogate metric is to make decision on the true north metric early, not to being inconclusive in every experiment we run. To leverage the full potential of surrogate metrics, we need to improve the comparison sensitivity. One way to achieve that is through variance reduction.

There are many approaches for variance reduction. One of the widely used approach in the online experiment community is CUPED [7] due to its simplicity and the level of variance reduction it can achieve. CUPED leverages experiment unit's pre-experiment covariates for variance reduction. Since the majority of our experiments discussed in this paper target members on LinkedIn, fortunately, we have a good set of member covariates. Specifically, for the job marketplace use case, we have seen that members who actively apply for jobs this week will have a high chance to continue actively applying for jobs the following week.

After performing CUPED variance reduction, we need to repeat the same adjustment on PCH to include the prediction error. As expected, the comparisons on adjusted PCH after variance reduction become more statistically significant. Out

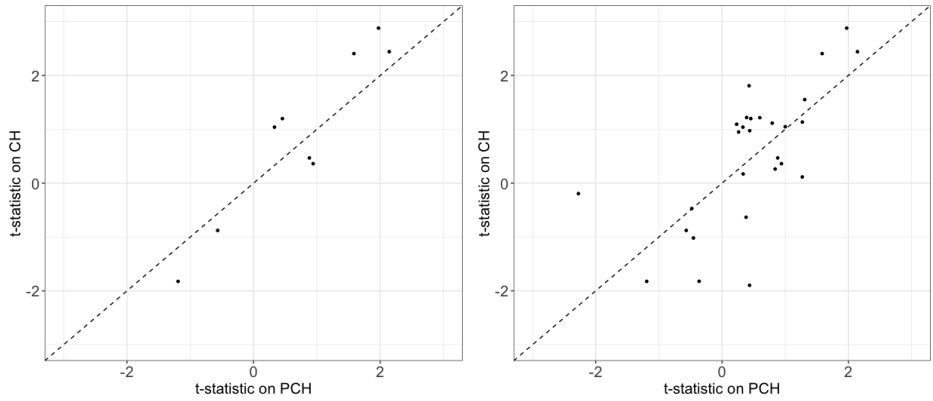

Fig. 7. Adjusted PCH t-Statistic vs. CH t-Statistic. Left: Experiments with Sample Size > 1M in Both Treatment and Control; Right: Experiments with Sample Size > 10K in Both Treatment and Control. Same Set of Experiments as Figure 5)

of the 203 experiments we studied, 10 are significant on adjusted PCH. Recall that before variance reduction, we have only 2 significant comparisons. For the 10 significant comparisons, we examined the relationship of adjusted-PCH vs. CH before and after variance reduction. Figure 8 plots t-statistics from adjusted PCH and CH before and after variance reduction. For most of the experiments, CUPED provides 30%-50% variance reduction. In addition, Figure 8 shows that PCH and CH move in the same direction for all experiments we validated. This provides much confidence in trusting PCH in experiment setting. Given that CH is not available for several months while PCH is available within a couple of days, using PCH in online experiments saves us several months in measuring the impact on the true-north metric (CH) and in making optimal ramp decisions.

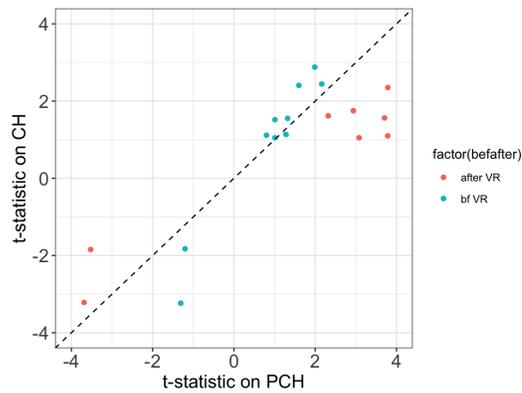

Fig. 8. t-Statistic from Adjusted PCH vs. CH. Before Variance Reduction in Cyan; After Variance Reduction in Red

### 5.5 Onboarding PCH onto Experimentation Platform for Decision Making

With PCH properly defined and widely adopted by engineering and product teams, we have onboarded this metric onto LinkedIn's Unified Metric Platform [29]. We then leverage the commonly used two-sample t-test to conduct the comparison between treatment group and control group after variance reduction and predictive error adjustment. Figure 9 shows PCH on LinkedIn experimentation platform. Before having PCH, users of the experiment platform rely on the treatment impacts from a set of metrics. Quite often, the treatment impacts may not align with each other in the same direction. In such scenarios, trade-off has to be made before ramping further. It is not at all principled how users make such trade-off. Having PCH as the success metric for decision has cleared up the confusion and encouraged unified and consistent ramp practice on all job marketplace experiments.

| Metric Name | % Change | p-value | Confidence Interval |
|---|---|---|---|
| Job Apply Predicted Confirmed Hire 6m | +0.84% | 0.0034 | [+0.28% , +1.40%] |

Fig. 9. PCH on LinkedIn Experimentation Platform

## 6 CONCLUSION AND FUTURE WORK

In this paper we discussed the use of surrogate metrics in experimentation setting, the practical challenges and our solutions. We first discussed how to pick a good surrogate metric with both statistical and practical considerations. We then dived into the experiment analysis challenges after picking a surrogate metric. After building the foundations, we switched to an important application of surrogate metric in LinkedIn job marketplace to demonstrate how we leverage surrogate metrics for quick decision making.

One interesting angel we have not studied in depth is how the surrogate metric predicts the true north with respect to the experiment duration. To study this dynamic, we need to have hold-out experiments where we can continue to observe the surrogate metric and the true north. With this information, we can make better decisions on the optimal experiment duration balancing surrogacy predictive power and experiment cost.

## ACKNOWLEDGMENTS

We thank our researchers and scientists in Data Science Applied Research and Hiring Marketplace Data Science for their feedback.